\documentclass[preprint,showpacs,preprintnumbers,amsmath,amssymb]{revtex4}
\usepackage{graphicx}

\begin{document}

\draft

\title{Theory of single photon transport in a coupled Semiconductor microcavity waveguide}
\author{Yin Zhong, Lei Tan\footnote{Corresponding author. Electronic address:
tanlei@lzu.edu.cn}, Li-wei Liu}
\address{Institute of Theoretical Physics, Lanzhou University, Lanzhou 730000, P. R. China}

\begin{abstract}
We investigate the coherent transport of a single photon in
coupled semiconductor microcavity waveguide,which can be
controlled by in-plane excitons in quantum well embedded in the
antinode of the electromagnetic field in one of the cavities. The
reflection coefficient and transmissivity for the single photon
propagating in this semiconductor waveguide are obtained. It is
shown that the effect of the exciton's decay plays an important
role in the transport properties of the single photon in this
microcavity waveguide if we refer to real systems.
\end{abstract}

\pacs{03.75.Mn, 75.40.Gb }

\maketitle

\section{Introduction}
Recent years, many theoretical and experimental researches have
been focused on semiconductor
microcavity\cite{Deng,Snoke,Tsintzos,Kasprzak,Khitrova}. Most of
these researches are devoted to the Bose-Einstein Condensation of
exciton polaritons which are quasi-two dimensional bosons in the
semiconductor microcavity and even BEC of polaritons in a
trap\cite{Kim,Balili,Nelsen,Berman}. But here we are interested in
the coherent transport of single photon in semiconductor
microcavity waveguide because of the  impressive result achieved
in one-dimensional resonator waveguide in optical coupled
cavities\cite{Zhou, Jung1,Jung2}. In Ref.\cite{Zhou}, the results
show that the controllable two-level system can behave as a
quantum switch for the coherent transport of a single photon which
may be a good candidate for quantum information processing where
controlling of coherent transport for a scattering photon by
tuning inner structure of the target is a hot
topic\cite{Zhou,Lu,Hu}.

A typical semiconductor microcavity consists of a planar
Fabry-Perot cavity sandwiched between so-called Bragg mirror,
which is a period structure composed of two semiconductor or
dielectric materials with different refractive indices and
containing an embedded quantum well(QW). And the semiconductor
microcavity waveguide(consists of many cavities)can be fabricated
by conventional manufacture technology of
semiconductor\cite{Kavokin}. Compared with the optical cavities,
the semiconductor microcavity waveguide has several XXXX
characteristics in application. Firstly, the practical temperature
of this semiconductor system can be several orders higher than the
countpart in optical microcavity waveguide in some point, and this
is quite important for future applications\cite{Senellart,
Weisbuch,Press}in quantum information processing. Secondly, the
excitons in the QW are Coulomb-corelated electron-hole pairs
characterised by discrete transition frequencies,which can be
treated as two-level systems\cite{Eastham}. This is very similar
with the two-level atom in optical cavity waveguide\cite{Zhou}. At
last, the excitons are stationary against quite high
temperature\cite{Tsintzos,Baumberg}. By contrast, the cold atoms
are only existing in very low
temperature\cite{Chu,Cohen-Tannoudji,Phillips}, and it is
cumbersome for people to prepare a cold atom initially and load it
into the cavity waveguide. Considering these reasons, studying the
the coherent transport  of a single photon in semiconductor
microcavity waveguide is very interesting. We will get reflection
coefficient and transmissivity for a single photon transporting in
the waveguide by solving eigen-equation of this coupled system
composed of single photon and exciton. as a matter of fact,we do
not add a pump laser to excite a real exciton, even though the
same function as setup in Zhou $et.al.$\cite{Zhou} can also be
realized. Which is duo to the reasons that, when no exciton is
excited, the state is chosen as low-energy level, and if one
exciton(this exciton is only excited by the photon propagate in
the waveguide)is excited, we call this state up-energy level, and
more excitons are not necessary. Furthermore the decay of excitons
has also been investigated for its impressive effect on scattering
if we refer to real systems, and this effect plays an important
role in the transport properties in this microcavity waveguide.

This paper is organized as followed. In Section $2$ we describe
the Hamiltonian for this system and derive an eigen-equation for
propagating of single photon in the waveguide. In section $3$, we
use simple wave function to obtain reflection coefficient and
transmissivity and analyze the parameters in these two
coefficients, in section $4$, we study the effect of decay for
excitons phenomenologically, finally we conclude in section $5$.

\section{The model for the coupled microcavity waveguide}
We describe the system as a single photon propagates in a
semiconductor microcavity waveguide(here the number of cavity is
infinite for we consider a ideal situation,however,in reality a
waveguide must have at least a input and a output,so the real one
may have several differences from our ideal system here)and
excitons transport in the plane of QW in one of cavities in the
waveguide. And we only restrict ourselves to a single photon
propagating state, this means initially there are no exciton in
the cavity and only after a photon is absorbed by the QW, an
exciton is created and then the exicton breaks down and a photon
is recreated but moves in the opposite direction to the incident
direction. As we know there is not a exiting Hamiltonian for this
system, so the first step to solve the problem is to model a
Hamiltonian where we must consider propagating of photon in
mirocavity waveguide and excitons in the plane of quantum well in
one of the cavity.In general, the widely used
Hamiltonian\cite{Berman} is given by
\begin{subequations}
\begin{equation}
H=H_{exc}+H_{ph}+H_{exc-ph} \label{eq-1}
\end{equation}
\begin{equation}
H_{exc}=\sum_{p}\varepsilon_{ex}(p)b^{\dag}_{p}b_{p}+\frac{1}{2A}\sum_{p,p^{'},q}U_{q}b^{\dag}_{p+q}b^{\dag}_{p^{'}-q}b_{p}b_{p^{'}}
\end{equation}
\begin{equation}
H_{ph}=\sum_{p}\varepsilon_{ph}(p)a^{\dag}_{p}a_{p}
\end{equation}
\begin{equation}
H_{exc-ph}=\Omega_{R}\sum_{p}a^{\dag}_{p}b_{p}+h.c.
\end{equation}
\end{subequations}
$H_{exc}$ is an excitonic Hamiltonian,$H_{ph}$ is a photonic
Hamiltonian and $H_{exc-ph}$ is a Hamiltonian which describes the
interaction between photons and excitons. Here, $a_{p}$ and
$a^{\dag}_{p}$ are photonic creation and annihilation operators,
$b_{p}$ and $b^{\dag}_{p}$ are excitonic creation and annihilation
operators which satisfy Boson commutation relation.
$\varepsilon_{ex}(p)=E_{band}-E_{binding}+\varepsilon_{0}(p)$ is
the energy for single exciton state,
$E_{binding}=Ry_{2}^{\ast}=\mu_{e-h}e^{4}/\hbar^{2}\epsilon$ is
the binding energy for interaction between 2D holes and
electrons,$\mu_{e-h}=m_{e}m_{h}/(m_{e}+m_{h})$ is the reduced
excitonic mass, $\epsilon$ is the dielectric constant, $e$ is the
charge of an electron,
$\varepsilon_{0}(p)=p^{2}/2M,M=m_{e}+m_{h}$, $m_{e},m_{h}$ is the
mass of electron and hole respectively. $A$ is the macroscopic
area which we refer to the excited area of
pump-beam\cite{Berman,West}. $U_{q}$ is the Fourier transform of
exciton-exciton pair repulsion potential. In real system, the
density of excitons is very low and this potential can be replaced
by an effective potential $U_{q}\simeq U_{0}\equiv
U=6e^{2}a_{2D}/\epsilon$,$a_{2D}=\hbar^{2}\epsilon/(2\mu_{e-h}e^{2})$
is 2D Bohr radius of excitons\cite{Ciuti,Ben-Tabou, Berman}.
$\varepsilon_{ph}(p)=\frac{c}{n}\sqrt{p^{2}+\hbar^{2}\pi_{2}L_{c}^{-2}}$
is photonic energy spectrum,$c$ is the speed of light in vacuum
and $n=\sqrt{\epsilon}$ is refractive index in the cavity.The
coupling of excitons and photons are attributed to the Rabi
frequency $\Omega_{R}=\frac{d}{2\pi
a_{2D}}\sqrt{\frac{N\varepsilon_{exc}}{\hbar^{2}n\lambda_{0}}}$,
which is proportional to the exciton oscillator strength(d) and to
the number of QWs embedded in the cavity with $\lambda_{0}$ the
resonant wavelength of the cavity.

This Hamiltonian is widely used to study exciton polaritons, which
are the elementary excitations of coupled systems composed of
matter(excitons)and light(photons)in strong coupling
regime\cite{Savona,Berman,Kavokin}. This Hamiltonian can be
modified for the application of considering the single photon
transport in the coupled semiconductor microcavity waveguide.
Firstly in the original Hamiltonian, the photons propagate in the
homogeneous space in the plane of QW, but in the third direction
electromagnetic field is confined in cavities. The momentum $p$ or
wave vector(if $\hbar=1$)is the in-plane one. Exciton-exciton
interaction and exciton-photon interaction conserve these in-plane
momentums. In this paper we want to analyze the problem of
scattering of a single photon in coupled semiconductor microcavity
waveguide, the effect of microcavity waveguide must be included in
the photonic Hamiltonian and the original photonic Hamiltonian can
be replaced by the following one\cite{Hu,Zhou,Lu}.
\begin{equation}
H_{ph}=\omega\sum_{j}a^{\dag}_{j}a_{j}-\xi(\sum_{j}a^{\dag}_{j}a_{j+1}+h.c).
\end{equation}
In this Hamiltonian, we only consider one mode whose incident
angle is zero toward to the plane of QW,  and  in-plane momentum
of this mode vanishes too. We use $a^{\dag}_{j}(a_{j})$to denote
the creation(annihilation) operator of the $j$th cavity. $\xi$ is
the nearest-neighbor evanescent coupling constant of intercavity,
and $\omega$ is the energy of photon in each cavity.

Secondly, the photon only couples with the excitons carrying the
same momentum as the them. This can be inferred from the
interaction Hamiltonian  $H_{exc-ph}$. we can get the new
Hamiltonian for coupled excitons and photons.
\begin{equation}
H_{exc-ph}=\Omega_{R}(a^{\dag}_{0}b_{0}+a_{0}b^{\dag}_{0})
\end{equation}
$a^{\dag}_{0}(a_{0})$is the photonic
creation(annihilation)operator in the zeroth cavity for we assume
QWs(or excitons) only locate in this cavity(plane),and
$b^{\dag}_{0}(b_{0})$creates(annihilates)excitons in the zeroth
cavity for momentum is vanished.$\Omega_{R}$is Rabi frequency for
this coupling. As what we have done in Hamiltonian $H_{exc-ph}$,
we are able to have a replaced Hamiltonian for excitons which only
stay in the QWs' plane in the zeroth cavity.
\begin{equation}
H_{exc}=\varepsilon_{ex}(0)b^{\dag}_{0}b_{0}+\frac{1}{2A}Ub^{\dag}_{0}b^{\dag}_{0}b_{0}b_{0}
\end{equation}
then a suitable Hamiltonian for our problem on scattering of a
single photon in coupled semiconductor microcavity waveguide can
be given as following:
\begin{equation}
H=H_{ph}+H_{exc}+H_{exc-ph}=\omega\sum_{j}a^{\dag}_{j}a_{j}-\xi\sum_{j}(a^{\dag}_{j}a_{j+1}+h.c.)
+\varepsilon_{ex}(0)b^{\dag}_{0}b_{0}+\frac{1}{2A}Ub^{\dag}_{0}b^{\dag}_{0}b_{0}b_{0}+\Omega_{R}(a^{\dag}_{0}b_{0}+a_{0}b^{\dag}_{0})
\end{equation}
In the following, we do not deal with it directly(mean field
theory,perturbation theory, Hubbard-Stratonovich transformation
and so on\cite{Altland}). Here we do is to find the scattering of
a single photon in this coupled cavity waveguide,so we consider an
eigen-equation for this physical process.  To get an
eigen-equation, we have to assume an eigen-vector and this can be
done as follows.We focus on $H_{ph}$,then let the waveguide be
infinite and make a Fourier transformation for
$a_{j}(a^{\dag}_{j})$.$a_{j}=\frac{1}{\sqrt{N}}\sum_{k}e^{ikj
l}a_{k}$\cite{Ashcroft} where $k$ is the wave vector of photons,
$N$ is the number of cavities in the waveguide and $l$ is the
distance between neighbor cavities. So, in momentum(wave vector)
space the photonic Hamiltonian can be expressed as
\begin{equation}
H_{ph}=\sum_{k}\Omega_{k}a^{\dag}_{k}a_{k}=\sum_{k}(\omega-2\xi\cos(kl))a^{\dag}_{k}a_{k}
\end{equation}
$\Omega_{k}=\omega-2\xi\cos(kl)$. For simplicity, we let $l=1$
below, so $\Omega_{k}=\omega-2\xi\cos(k)$. Now we assume the
stationary eigen-vector\cite{ Hu,Zhou} is
\begin{equation}
|\Omega_{k}\rangle=\sum_{j}a^{\dag}_{k}\mu_{k}(j)|0\rangle_{c}|0\rangle_{exc}+\mu^{'}|0\rangle_{c}|1\rangle_{exc}
\end{equation}

$|0\rangle_{c}$ is the vacuum state of photons,
$|0\rangle_{exc}(|1\rangle_{exc})$ is the state of zero momentum
for excitons in QWs, $\mu_{k}(j)$ and $\mu^{'}$ are the
probability amplitudes of the excitons in different
particle-number state. Then let us make an eigen-equation
$H|\Omega_{k}\rangle=\Omega_{k}|\Omega_{k}\rangle$and we can get a
discrete scattering equation
\begin{equation}
(\omega-\Omega_{k}+\frac{\Omega^{2}_{R}}{\Omega_{k}-\varepsilon_{ex}(0)}\delta_{j,0})\mu_{k}(j)=\xi\{\mu_{k}(j+1)+\mu_{k}(j-1)\}
\end{equation}

We should notice that the eigen-vector only involve two sub-state.
One is only a photon in the waveguide and the other is only one
exciton excited. Actually, we may go beyond this regime and we add
one or more excitons to the QWs' plane, but the theory we give
here fail because more than one photon can exist for excited
emission of excitons and this rule out the assumption that only
one photon propagate in the waveguide. This is because that if no
more than one exciton exist, exciton-exciton interaction can be
neglected, so this is not a many-body problem and we can get
analytical result for this problem.

\section{Reflection coefficient and Transmissivity}
With the discrete scattering equation,and considering its
similarity to standard quantum mechanics scattering problem,we
assume the solution of this equation is
\begin{equation}
\mu_{k}(j)=\{e^{ikj}+re^{-ikj}\}(j<0),\nonumber
\end{equation}
\begin{equation}
\mu_{k}(j)=se^{ikj}( j>0)
\end{equation}
$r,s$ are the reflection and transmission amplitudes. Then we
insert assumed solution above to eq.(8) and get
\begin{equation}
r=\frac{\Omega_{R}^{2}}{(2i\xi\sin(k))(\omega-\varepsilon_{ex}(0)-2\xi\cos(k))-\Omega_{R}^{2}}
\end{equation}
\begin{equation}
s=r+1
\end{equation}
the reflection coefficient is $R=|r|^{2}$ and transmissivity is
$T=1-R$
\begin{equation}
R=\frac{1}{1+4(\xi/\Omega_{R})^{2}\sin^{2}(k)(\omega/\Omega_{R}-\varepsilon_{ex}(0)/\Omega_{R}-2\xi/\Omega_{R}\cos(k))^{2}}
\end{equation}
\begin{equation}
T=\frac{4(\xi/\Omega_{R})^{2}\sin^{2}(k)(\omega/\Omega_{R}-\varepsilon_{ex}(0)/\Omega_{R}-2\xi/\Omega_{R}\cos(k))^{2}}{1+4(\xi/\Omega_{R})^{2}\sin^{2}(k)(\omega/\Omega_{R}-\varepsilon_{ex}(0)/\Omega_{R}-2\xi/\Omega_{R}\cos(k))^{2}}
\end{equation}
Figs. $1$, $2$ show the reflection and transmission coefficients
as a function of momentum $k$,respectively. Fig.$3$ combines both
of coefficients. From these results, we see the system can be
considered as a semiconductor switch where we might control
excitons' parameters so as to make a giant influence on the
photon's propagating in this semiconductor microcavity waveguide.
The corresponding system has been studied in optical system(or
called cavity quantum electrodynamics)by using a two-level atom as
a controller recently\cite{Zhou}. And our system is a conventional
solid state system,this may be achieved at several
kelvin\cite{Senellart, Weisbuch,Press}.

Apparently, this is a good result because we have a
parameter-dependence reflection coefficient and transmissivity for
all existing quantities, however, in fact we do not know these
parameters directly. Instead, we change this into another form
\begin{equation}
R(\delta)=\frac{1}{1+4(\xi/\Omega_{R})^{2}\delta^{2}[1-(\frac{\delta-\omega+\varepsilon_{ex}(0)}{2\xi})^{2}]}
\end{equation}
\begin{equation}
T(\delta)=\frac{4(\xi/\Omega_{R})^{2}\delta^{2}[1-(\frac{\delta-\omega+\varepsilon_{ex}(0)}{2\xi})^{2}]}{1+4(\xi/\Omega_{R})^{2}\delta^{2}[1-(\frac{\delta-\omega+\varepsilon_{ex}(0)}{2\xi})^{2}]}
\end{equation}
and parameters in this form can be easily measured from
photoluminescence experiments\cite{Kavokin,Khalifa,Bajoni}. Here
$\delta=\omega-\varepsilon_{ex}(0)-2\xi\cos(k)$ is detuning of
photons and excitons(It is quite different from conventional
definition and we will discuss this below). Figs. 4, 5 show result
of eq.(14) and eq.(15) where we find detuning is a quite important
parameter and an easily controllable quantity in this
waveguide\cite{Kavokin}. Fig.6 shows two coefficients of
reflection and transmission changing with different detuning.
Generally, in semiconductor microcavity detuning(there the
detuning is defined as
$\delta=\varepsilon_{ph}(k=0)-\varepsilon_{ex}(k=0)$), which is
smaller than Rabi frequency($\Omega_{R}$), and this means that
$\omega-\varepsilon_{ex}(0)$ is smaller than $\Omega_{R}$ in our
paper\cite{Kavokin}, so in Figs.1, 2, 3, 4, 5 and 6, we choose
$\omega=\varepsilon_{ex}(0)=100\Omega_{R}$, $\xi=2\Omega_{R}$ for
 the energy of photons or excitons, which is
the order of $eV$ while $\Omega_{R}$ is two orders lower than
them\cite{Tsintzos,Kasprzak}in semiconductor microcavity.

\section{Effect of decay for excitons}
In solid state systems, excitons are elementary excitations of
electrons and holes in semiconductor, and they will decay due to
interactions with other particles, for example phonons, free
electrons or holes. For simplicity, we can let
$\varepsilon^{'}_{ex}(0)=\varepsilon_{ex}(0)-i\Gamma$ and $\Gamma$
is a phenomenological decay rate that accounts for interactions
mentioned above(but decay of excitons due to phonons which are
excited by temperature is essential for no pumping of other
particles).  Then we can get the  new reflection and transmission
coefficients by using this replacement.
\begin{equation}
R=\frac{1}{(1+\frac{2\xi\sin(k)\Gamma}{\Omega_{R}^{2}})^{2}+4(\frac{\xi}{\Omega_{R}})^{2}\sin^{2}(k)(\omega/\Omega_{R}-\varepsilon_{ex}(0)/\Omega_{R}-2\xi/\Omega_{R}\cos(k))^{2}}
\end{equation}
But transmission coefficient $T\neq1-R$ and
$T=|s|^{2}=|1+r|^{2}=1+r+r^{\ast}+R$. This is quite common for we
include dissipation in our model system,  and particles(here the
excitons) may be absorbed by the environment around them. Exciton
is excited by the incident photon and conservation for the number
of excitation will fail if exciton decays(absorbed by
environment). To compare with non-dissipation reflection
coefficient,we have to estimate the magnitude of $\Gamma$. Here,
we know the intrinsic exciton lifetime is about 100ps(for
temperature is 4K)\cite{Berman}, and we estimate
$\Gamma\sim0.01\Omega_{R}$. This result are shown in Fig.7, and we
see the central peak is nearly unchanged, but intensity of the
second peak on the right is not unity(about 90 percent)due to the
dissipation(but in optical waveguide proposed by Lan Zhou
$et.al.$\cite{Zhou}, we use conventional parameters\cite{Maunz}to
estimate the effect of dissipation and we find this value is about
60 percent impressively). However, the second peak on the left do
not show this effect, because initially we only consider a photon
propagate from left to the right of the waveguide what results in
this asymmetry.

One may ask the reason why we here do not consider the decay of
photons, this is because in the microcavity, if photons escape
from the cavity, we call this decay, but here do not exist this
one and it has been included in the hopping term of photonic
Hamiltonian($H_{ph}$), and other channels of decay are so weak in
comparison with decay of excitons\cite{Kavokin}.

\section{ Summary}
In this paper we model a Hamiltonian for transport of a single
photon in semiconductor microcavity waveguide, then we deal with
eigen-equation of this scattering and get reflection and
transmission coefficients from this equation. Parameters in two
coefficient can be easily measured using usual solid state
method(photoluminescence experiments). Two reflection coefficients
have been derived and discussed. We show that the photon in
coupled semiconductor microcavity waveguide can be controlled by
in-plane excitons in quantum well. And we also investigate the
effect of decay for excitons and this shows reduction of
reflection intensity for various momentum although the reduction
is about $10$ percent at 4K. For applications, the higher the
practical temperature, the more attractive the setup. And this is
the reason we are interested in this semiconductor system. We
think it is also a candidate for quantum information process
mentioned in the introduction of this paper.

But we here consider the waveguide as an ideal one, in other
words, the number of cavity in the waveguide is infinite and in
reality this has to be modified, so a more carefully consideration
should solve this problem and effect of temperature also has to be
more carefully deal with for this will strongly enforce the
interaction between excitons and phonons\cite{Kavokin}and we here
only include this effect in the energy of excitons
phenomenologically. After all, we analyze the scattering of a
single photon in semiconductor microcavity waveguide and we
propose experiments may be done in semiconductor systems for its
high practical temperature, easily fabrication and
measurement\cite{Kavokin}.

\section{Acknowledgements}

This work was partly supported by the National Natural Science
Foundation of China under Grant No.$10704031$, the National
Natural Science Foundation of China for Fostering Talents in Basic
Research under Grant No.$J0730314$, the Fundamental Research Fund
for Physics and Mathematics of Lanzhou University under Grant No.
Lzu$05001$, and the Nature Science Foundation of Gansu under Grant
No. 3ZS061-A25-035.

\newpage
\begin{figure}
\centering
\includegraphics[width=5.5in,height=4.5in]{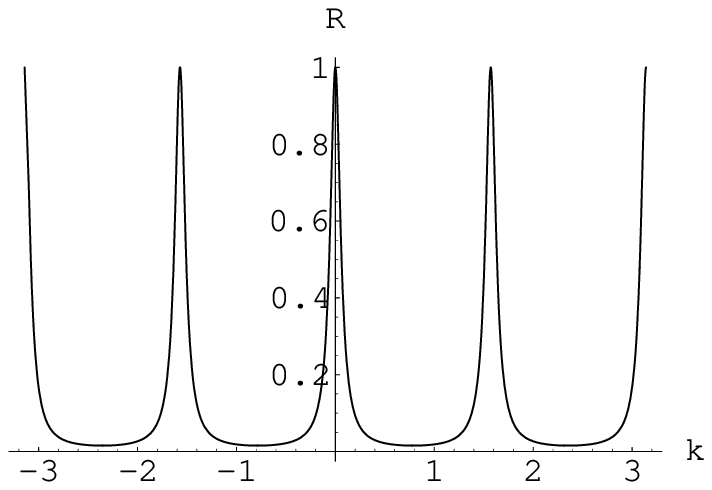}
\caption{Reflection coefficient $R$ as a function of momentum $k$
with $\xi=2\Omega_{R}$,$\omega=\varepsilon_{ex}(0)=100\Omega_{R}$}
\end{figure}

\newpage
\begin{figure}
\centering
\includegraphics[width=5.5in,height=4.5in]{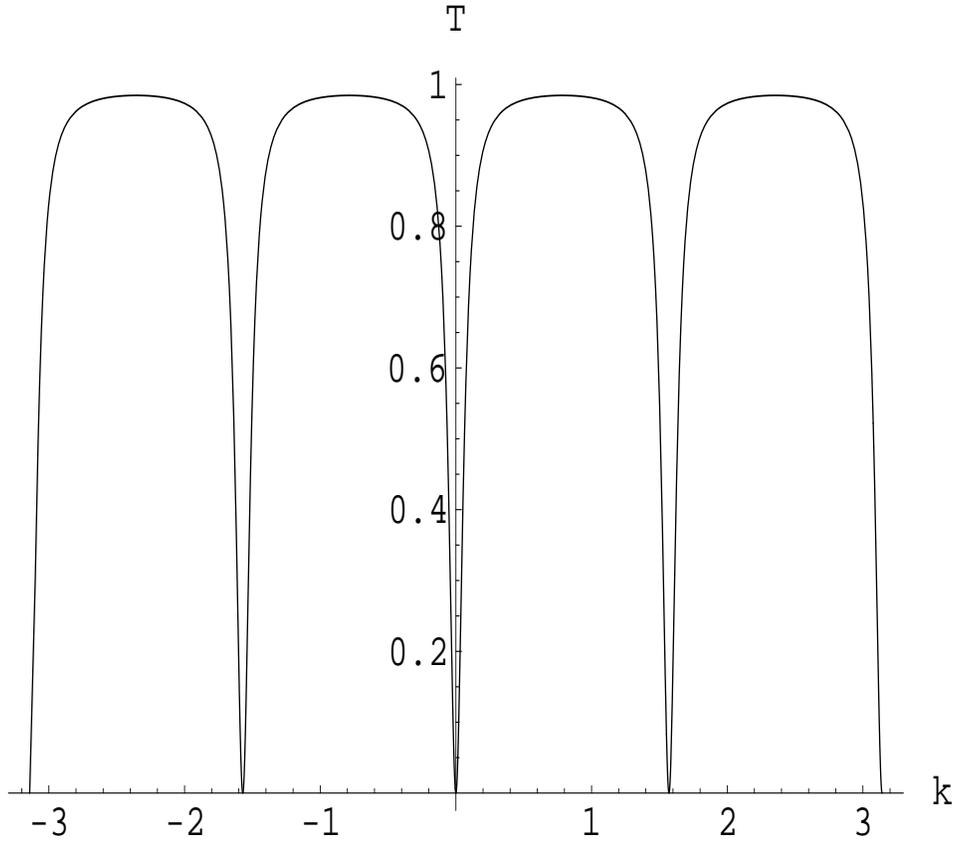}
\caption{Transmission coefficient $T$ as a function of momentum
$k$ with
$\xi=2\Omega_{R}$,$\omega=\varepsilon_{ex}(0)=100\Omega_{R}$}
\end{figure}

\newpage
\begin{figure}
\centering
\includegraphics[width=5.5in,height=4.5in]{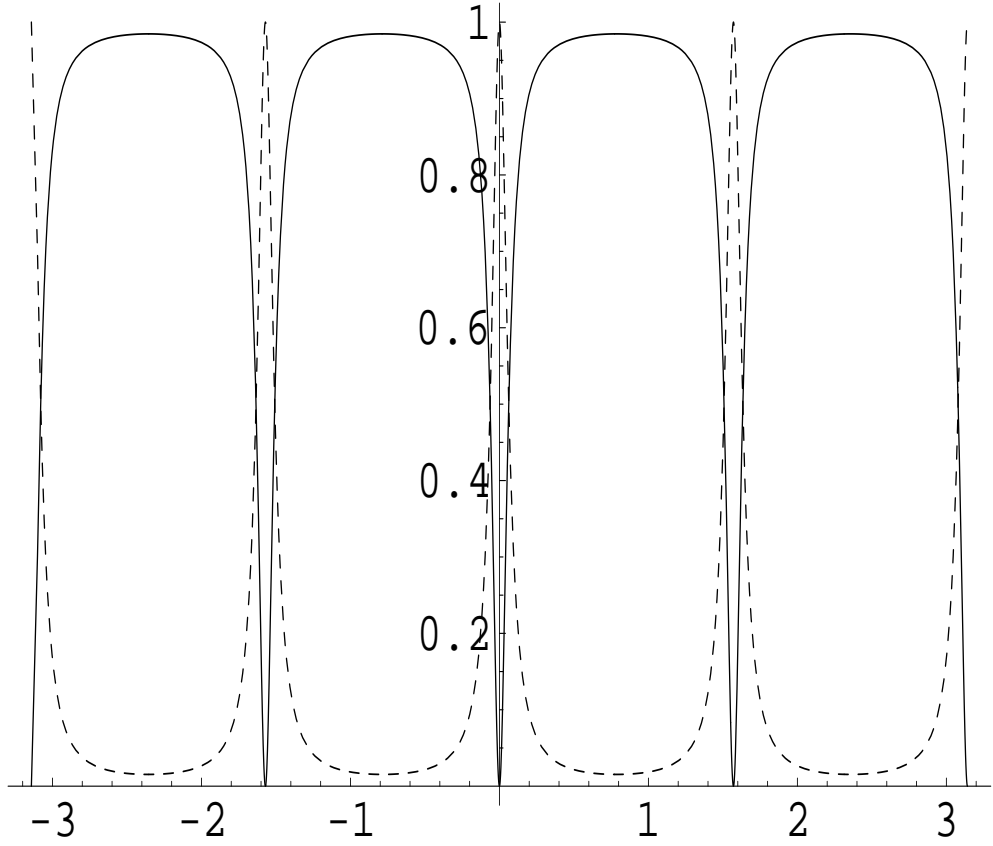}
\caption{Reflection(dash line) and transmission(solid line)
coefficients as a function of momentum $k$ with
$\xi=2\Omega_{R}$,$\omega=\varepsilon_{ex}(0)=100\Omega_{R}$}
\end{figure}

\newpage
\begin{figure}
\centering
\includegraphics[width=5.5in,height=4.5in]{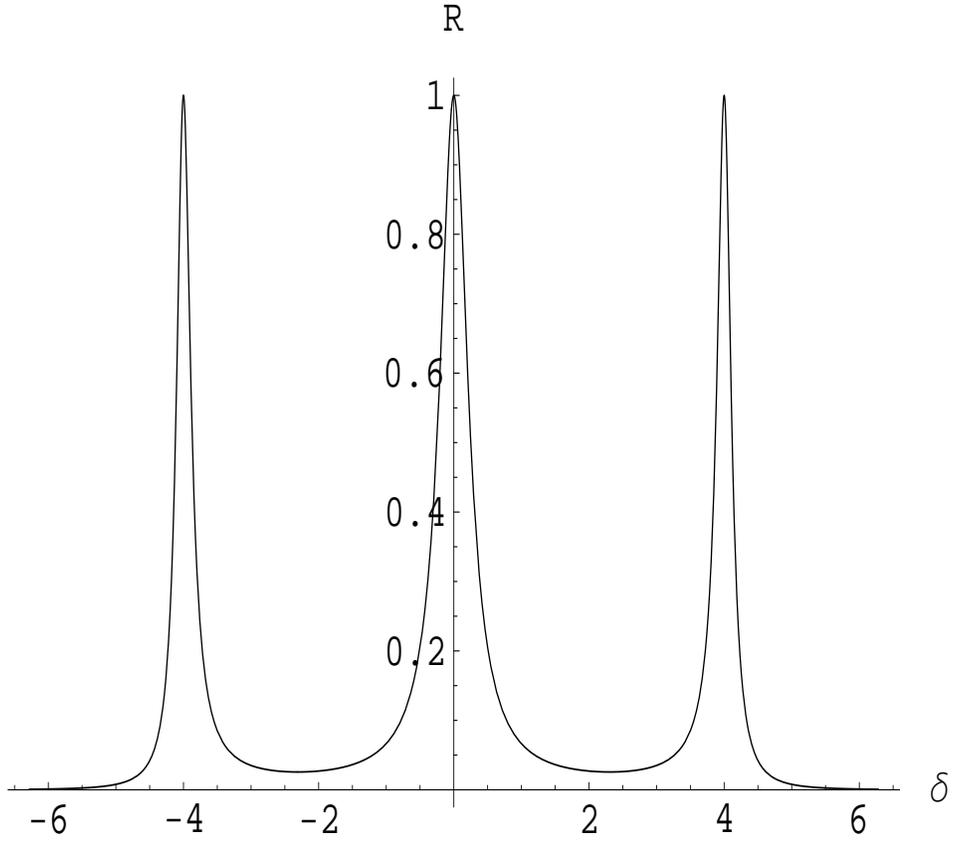}
\caption{Reflection coefficient $R$ as a function of detuning
$\delta$ with
$\xi=2\Omega_{R}$,$\omega=\varepsilon_{ex}(0)=100\Omega_{R}$}
\end{figure}

\newpage
\begin{figure}
\centering
\includegraphics[width=5.5in,height=4.5in]{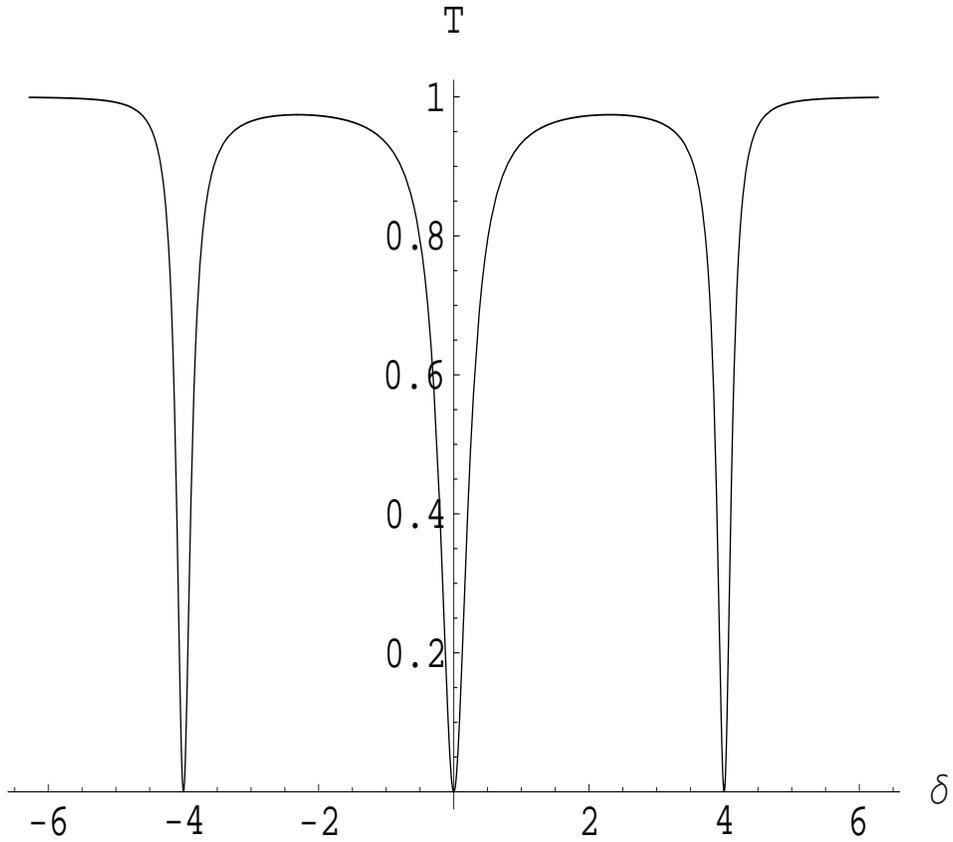}
\caption{Transmission coefficient $T$ as a function of detuning
$\delta$ with
$\xi=2\Omega_{R}$,$\omega=\varepsilon_{ex}(0)=100\Omega_{R}$}
\end{figure}

\newpage
\begin{figure}
\centering
\includegraphics[width=5.5in,height=4.5in]{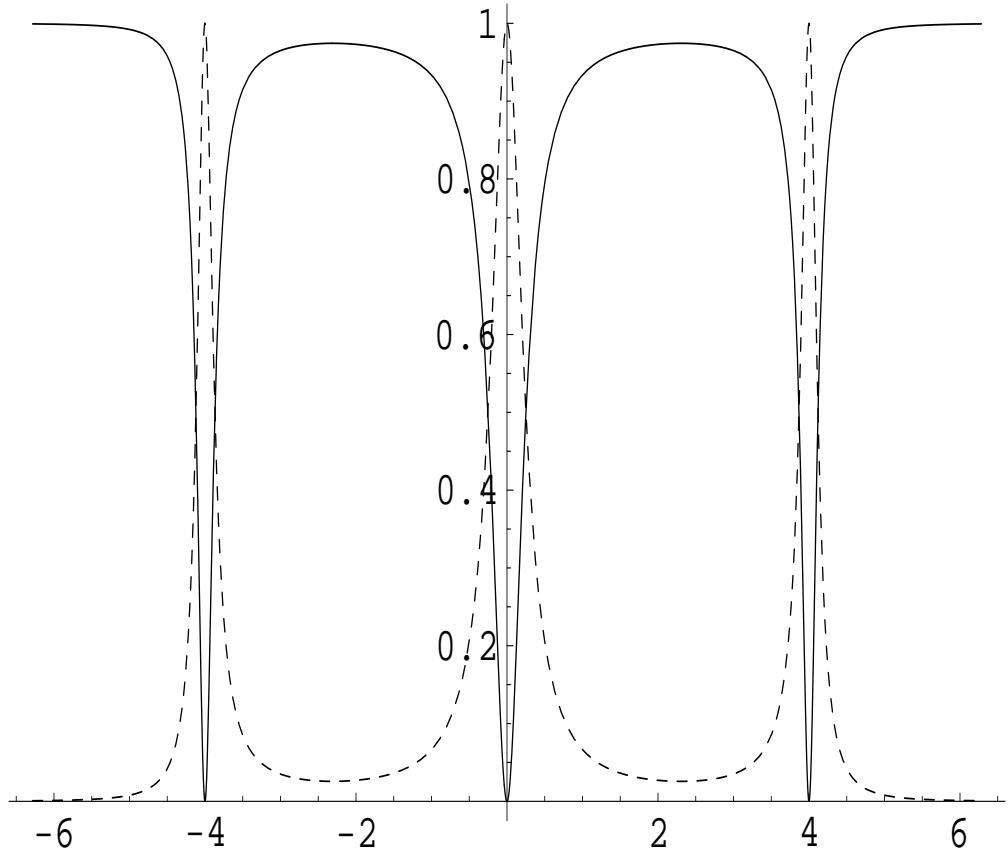}
\caption{Reflection(dash line) and transmission(solid line)
coefficients as a function of detuning $\delta$ with
$\xi=2\Omega_{R}$,$\omega=\varepsilon_{ex}(0)=100\Omega_{R}$}
\end{figure}

\newpage
\begin{figure}
\centering
\includegraphics[width=5.5in,height=4.5in]{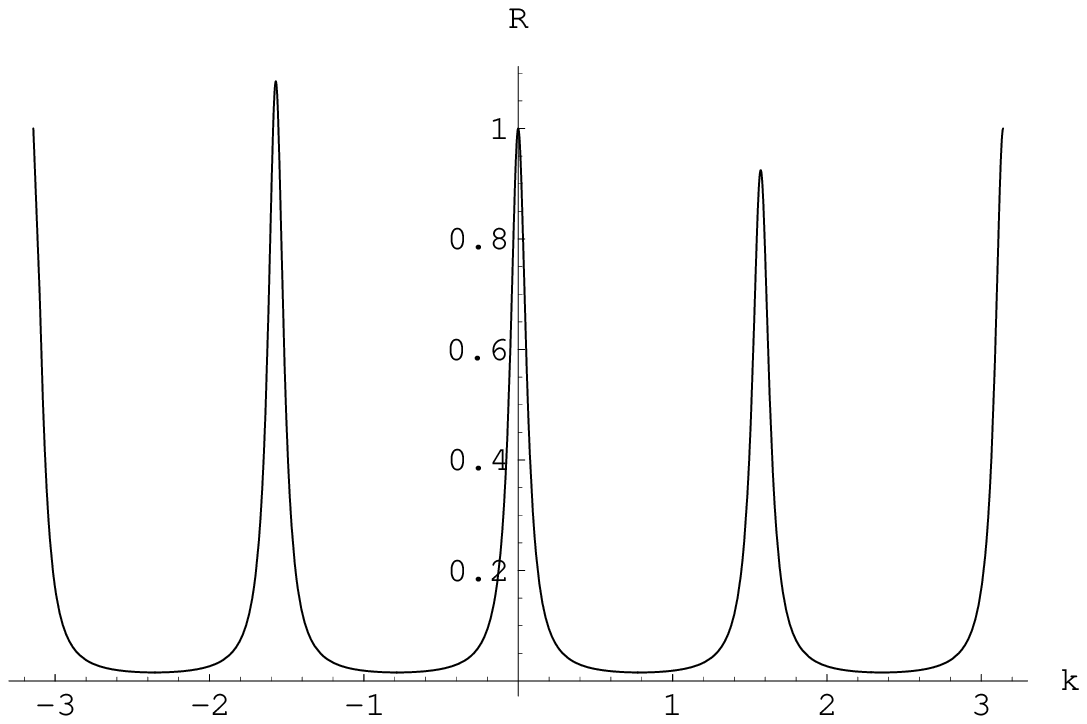}
\caption{Reflection coefficient $R$ as a function of momentum $k$
with $\xi=2\Omega_{R}$,$\omega=\varepsilon_{ex}(0)=100\Omega_{R}$
and $\Gamma=0.01\Omega_{R}$ }
\end{figure}
\end{document}